\documentclass[%
pre, aps, twocolumn
]{revtex4-1}
\usepackage{graphicx}
\usepackage{dcolumn}
\usepackage{color}
\usepackage{bm}
\usepackage{mathtools}

\begin{document}

\title{Severe population collapses and species extinctions in multi-host epidemic dynamics}

\author{Sergei Maslov}
\thanks{Email: maslov@illinois.edu}
\affiliation{Department of Bioengineering and Carl R. Woese 
Institute for Genomic Biology, University of Illinois at 
Urbana-Champaign, Urbana, IL 61801, USA}

\author{Kim Sneppen}
\thanks{Email: ksneppen@gmail.com}
\affiliation{Center for Models of Life, 
Niels Bohr Institute, University of Copenhagen, 
2100 Copenhagen, Denmark}

\begin{abstract} 
Most infectious diseases including more than half of known human 
pathogens are not restricted to just one host, yet much of 
the mathematical modeling of infections has been limited to a 
single species. We investigate consequences of a 
single epidemic propagating in multiple species 
and compare and contrast it with the endemic steady state of the disease. 
We use the two-species Susceptible-Infected-Recovered (SIR) model 
to calculate the severity of post-epidemic collapses in populations 
of two host  species  as a function of their initial population sizes, the times 
individuals remain infectious, and the matrix of infection rates. 
We derive the criteria for a very large, extinction-level, 
population collapse in one or both of the species. 
The main conclusion of our study is that a single 
epidemic could drive a species with high mortality rate 
to local or even global extinction provided that it is 
co-infected with an abundant species. 
Such collapse-driven extinctions depend on
factors different than those in the endemic 
steady state of the disease.
\end{abstract}

\maketitle


\section*{Introduction}

Models of pathogen dynamics for the most part include only a single 
host  species  \cite{lloyd-smith} in spite of the fact that pathogens typically
infect multiple species. For example, more than half of human pathogens 
are known to be shared with at least one animal species
\cite{woolhouse2005,wolfe2007}.
Famous examples of diseases with multiple host
 species 
include cholera ({\it Vibrio cholerae}) commensal in a number zooplankton  species 
\cite{colwell,vezzulli}, bubonic plague (Yersinia pestis) 
co-infecting and spreading between humans and rats \cite{dyer1978}, 
and more recently the avian influenza virus \cite{liu2005}.
The steady state of dynamical equations where 
multiple hosts are infected by the same pathogen
was previously considered 
by Dobson \cite{dobson2004}.
This important study addressed the interplay between the diversity of hosts 
and the stability of disease's endemic state. 
In our study  we chose to focus on the transient (as opposed to the steady state) 
dynamics of a single epidemic 
as it is spreading in several species. 
While our mathematical formalism can be 
easily generalized to an arbitrary number of species, our main results 
can be already demonstrated for just two species. In what follows we 
use only this simpler two-species model. 

The dynamic of a single epidemic of a 
disease is often described in terms of the 
SIR (Susceptible -- Infected -- Removed) model \cite{kermack1933} and its 
variants: 
{\color{black} $dS/dt=-\beta S \cdot I$,
$dI/dt=\beta S \cdot I-\gamma I$ and $dD/dt=\gamma I$.} 
In this model individual members of the population 
susceptible to disease ($S$) become 
infected ($I$), and are subsequently removed from the pool 
spreading the disease due to either their death ($D$) or newly 
acquired immunity. While from the 
mathematical perspective there is no difference between death 
and complete resistance to disease, only the former results 
in population collapses that are the focus of our study. 
A well known property of the single species 
SIR model \cite{kermack1933} is that in the course of
the first epidemic the population of the host species drops 
to a much lower level than its ultimate steady state 
population in the endemic state of the diseases. When this population 
collapse is especially severe the host species is vulnerable 
to either local or even 
complete extinction. Such 
collapses along with extinction events 
triggered by them are the main focus of our study. 

The mass-action equations describing 
the dynamics of transitions between the three states of the SIR model can be described by 
a single key parameter, $R_{0}$ (equal to $\beta S(0)/\gamma$ in 
the notation used above), 
called the basic reproduction number or the epidemiological 
threshold. It is defined as the number of new 
infections caused by each infected individual
at the very start of the epidemic when the density of 
susceptible individuals is still close to $S(0)$ - its 
value at the start of the epidemic. Thus for $R_{0}>1$ the infection started by a very 
small number of infected individuals will (at least initially) exponentially amplify
and ultimately reduce the size of the susceptible population. 
In the opposite 
case $R_{0}<1$ the initial infection will quickly fizzle out 
and the population size will remain virtually unchanged. 
As the epidemic spreads, 
the number of susceptible targets declines, ultimately 
leaving $S(\text{collapse})$ survivors. 
For $R_{0} \gg 1$ the population collapse is given by the 
exponential function of $R_{0}$: 
{$S(\text{collapse} ) \simeq S(0)\exp[-R_{0}])$ for $0<I(0)\ll S(0)$}. 
{\color{black} The exponential decline in the number of these survivors of an epidemic 
as a function of $R_{0}$ can be derived through eliminating the non-linear term in the 
SIR model by measuring time in units of the number of deaths: 
$dS/dD=- (\beta /\gamma ) S$. Thus $S(t)=S(0)\exp[-\beta D(t)/\gamma]$ 
{\color{black} leading to the final number of survivors
after the epidemic ran its course} $S(\text{collapse})=S(0)\exp[-\beta D(\text{collapse})/\gamma]=
S(0)\exp[-\beta (S(0)-S(\text{collapse}))/\gamma] \simeq S(0)\exp[-\beta S(0)/\gamma]=S(0)\exp[-R_{0}]$.

{\color{black} It is illustrative to compare the size of the population after a collapse with its steady state value in the endemic state of the disease. Such endemic state requires a constant source of susceptible individuals which is traditionally realized by adding a small birth term to the SIR model (see e.g. Ref. \cite{dobson2004}).
The collapse $S(0) \to S(0)/\gamma]=S(0)\exp[-R_{0}]$ dramatically overshoots the 
endemic steady state}
population density $S(\text{steady state})=S(0)/R_{0}$
In the endemic state of the disease each infected individual transmits it to exactly one other susceptible individual thereby keeping a permanent infection going without exponential expansion or decay. Hence, 
$1=S(\text{steady state}) \beta/\gamma=
S(\text{steady state})R_{0}/S(0)$. 

A classic example of a pathogen-host ecosystem overshooting 
its steady state immediately after the first epidemic 
can be found e.g. in the experiments carried out in Ref. \cite{levin1977}: 
when a new phage was introduced into a bacterial population dominated by susceptible 
strains resulted in a bacterial population drop by roughly 5 orders of 
magnitude followed by a slow recovery to the steady state which is 
only one order of magnitude lower than the population at the start of the experiment.
Similar contrast between the initial population collapse is possible for 
epidemics of airborne diseases such 
as measles or small pox where 
$R_{0}$ could exceed 10 in an immunological naive population.
{\color{black} While measles or small pox do not always kill 
infected individuals, if a 
similarly contagious disease that is 
100\% lethal to its hosts was to emerge}, the initial epidemic-induced collapse 
$\exp(-R_{0})=\exp(-10) \sim 5 \cdot 10^{-5}$ would reduce 
host's population to much below its long-term steady state 
level of $1/R_{0} \sim 1/10$ achieved when (or if) such 
disease would become endemic. 
One expects a local extinction of the species if the
population of survivors after the epidemic,  
$S(\text{collapse}) \simeq S(0) \exp(-R_{0})$, 
drops below one individual. 

In this paper we model a single epidemic of a disease 
infecting multiple host species and investigate 
how its transient dynamics can result in a severe 
collapse or even local extinction of either of these species.
Such a scenario is realistic because epidemics routinely
spill over to other species, that is to say, diseases 
transiently or permanently transverse species boundaries.
For example, 
several Ebola epidemics in wild gorilla groups in 
central Africa happened between 2002 and 2003 resulted in 
90\%-95\% reduction in gorilla populations \cite{bermejo2006}. 
Such local near-extinction collapses 
have been blamed on ongoing spillover of the Ebola 
virus from its reservoir host, 
fruit eating bats, subsequently amplified by 
ape-to-ape virus transmission \cite{leroy2005}. 

\section*{Methods}
The two-host SIR model describes the 
disease propagation in species 1 and 2  
via the following system of ODEs:
\begin{eqnarray}
\frac{dS_1}{dt} & = & - \beta_{11} S_1 \cdot  I_1 - \beta_{12} S_1 \cdot I_2  
\label{succ}
\\
\nonumber
\frac{dI_1}{dt} & = & 
\beta_{11} S_1 \cdot I_1 +\beta_{12} S_1 \cdot I_2 - \gamma_1 I_1\\
\nonumber
\frac{dD_1}{dt} & = & \gamma_1 I_1 \\
\nonumber
\frac{dS_2}{dt} & = & - \beta_{21} S_2 \cdot  I_1 - \beta_{22} S_2 \cdot I_2  
\\
\nonumber
\frac{dI_2}{dt} & = & 
\beta_{21} S_2 \cdot I_1 +\beta_{22} S_2 \cdot I_2 - \gamma_2 I_2\\
\nonumber
\frac{dD_2}{dt} & = & \gamma_2 I_2
\end{eqnarray}
Here we assume density-dependent transmission 
mechanism characteristic of non-sexually or 
vector-transmitted diseases in well-mixed populations. 
We use the traditional notation 
\cite{dobson2004} where $\beta_{ij}$ is the matrix of 
transmission rates from species $j$ to species $i$. 

The SIR equations remain the same if instead of dying 
infected individuals recover with full immunity. However, since 
the focus of our study 
is on population collapses we choose to interpret $\gamma_i$ as the 
death rate of infected individuals of 
the species $i$ (see Discussion for a more general case including both 
death and recovery). $S_i$, $I_i$ and $D_i$ refer to population densities 
of susceptible, infected, and dead individuals in each of two species correspondingly. 
Our model describes the time course of a single epidemic 
during which we ignore 
births of new susceptible
individuals. 
This approximation is justified when the time from infection to
death
is fast compared to other timescales in the system.

The epidemic is initiated at time $t=0$ with a very small numbers of infected hosts 
in either one or both species: $I_1(0)\ll S_1(0)$, and $I_2(0) \ll S_2(0)$. 
In this limit the resulting population dynamics is independent of the exact values of $I_1(0)$ and $I_2(0)$.

\begin{figure}
\includegraphics[angle=0,width=1.0\columnwidth]{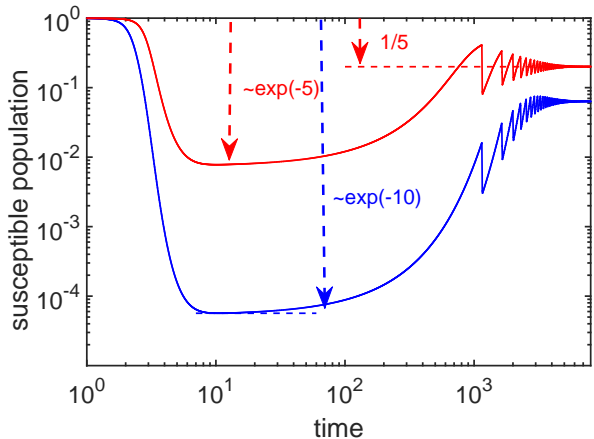}
\caption{Simulation of Eqs. \ref{succ} for populations of two species 
$S_1(t)$ (blue) and $S_2(t)$ (red) susceptible to infection by a single
pathogen with (cross)infection rates $\beta_{11}=\beta_{12}=\beta_{22}=5$, 
$\beta_{21}=0$, and $\gamma_1=\gamma_2=1$. 
At the start of
the simulation $S_1(0)=S_2(0)=1$ and $I_1(0)=0, I_2(0)=10^{-6}$.
In order for equations to have an endemic steady state we
added small birth and natural death terms as described in the text.
Right and left red arrows respectively point to 
the predicted steady state of the second population
$S_2(\text{steady state})=1/5$, and its population 
immediately after the first epidemic 
$S_2(\text{collapse}) \simeq \exp(-5)$.
The blue arrow highlights a much more severe post-epidemic 
collapse of the first species: 
$S_1(\text{collapse}) \simeq \exp(-10)$.
}
\label{fig0}
\end{figure}

\section*{Results}
{\color{black} We numerically simulated the time dynamics of Eqs. \ref{succ}, {\color{black} see Fig. 1.} 
To compare the results of a single epidemic to the endemic state of the disease 
we added a small birth term with saturation 
given by $0.01 \cdot S_i(t) \cdot (1-S_i(t))$ 
to the right hand side of the equations for $dS_i(t)/dt$. 
We also added even smaller natural (non-disease related) 
death term $-0.0001 \cdot S_i(t)$ to equations
for $dS_i(t)/dt$ and a similar death term  
$-0.0001 \cdot I_i(t)$ to equations for $dI_i(t)/dt$. 
This term ensures the flow of newly born susceptible 
individuals without affecting much 
either the population collapse after the first epidemic 
nor the long-term steady state of the system.
{\color{black} We then start our simulations at a 
pre-epidemic susceptible population $S_i(0)=0.99\sim 1$.}
The general steady state analysis of these equations has been 
carried out by Dobson \cite{dobson2004}.
The birth term used in our study differs slightly from that used 
in Ref. \cite{dobson2004} 
as we assume that infected individuals are infertile. 
When the growth rate is small (i.e. $\ll \gamma, \beta$) 
these choices do not significantly influence the collapse ratio
(data not shown).
In our interpretation of the SIR model 
the "removed" individuals are dead and thus 
(naturally) not included in the birth term.
This would change if infected individuals 
recover with a full immunity and are capable of 
giving birth \cite{dobson2004}.

We first consider a simple scenario
when the transmission is unidirectional $2 \to 1$. 
In this case an epidemic started in the 
species 2 would spill over to the species 1 and cause its 
population to collapse but not vice versa.
A case study is presented in Fig. \ref{fig0} where we 
plot time-courses of susceptible populations 
$S_1(t)$ (blue) and $S_2(t)$ (red)
defined by equations \ref{succ} with 
$\beta_{11}=\beta_{12}=\beta_{22}=5$, 
$\beta_{21}=0$, and $\gamma_1=\gamma_2=1$.
} 
For the case explored in Fig. \ref{fig0} the population dynamics of 
the second species is independent of the first one. Thus, like in a single species case outlined before, its epidemics is characterized by the basic reproduction number $S(0)\beta_{22}/\gamma_{2}=5$. In the endemic steady state its population 
is expected to be close to $S_2(\text{steady state})=1/5$ 
(the red dashed line in Fig. \ref{fig0}), while 
its initial post-epidemic collapse population to be 
approximately equal to  $S_2(\text{collapse})=\exp(-5)$ 
(the red arrow on the left of 
Fig. \ref{fig0}). 
{\color{black} Noticeably the species 1 (shown with blue) positioned 
``downstream" of the epidemics in the species 2, is exposed
to a much worse disease outbreak than the species 2. Its population collapses 
down to $S_1(\text{collapse})=\exp(-10) \ll S_2(\text{collapse})$. 
This amplification of outbreaks in two- or multi- host epidemics can be
described by the general theoretical framework described below.}

{\color{black} The equations 
\ref{succ} include both the unidirectional case discussed above, and the possibility that there is cross-infections in both directions. The equations can be solved by introducing two ``composite death toll'' 
variables $\tilde{D_1}=\beta_{11} D_1/\gamma_1+\beta_{12} D_2/\gamma_2$
and $\tilde{D_2}=\beta_{21} D_1/\gamma_1+\beta_{22} D_2/\gamma_2$. 
When these variables are used instead of time for each of two species, 
their susceptible populations follow 
a simple exponential decay 
$dS_1/d\tilde{D_1}=-S_1$ and $dS_2/d\tilde{D_2}=-S_2$ ending 
at their new post-collapse densities given by 
$S_1 (\text{collapse})=S_1(0)\exp[-\tilde{D_1}(\text{collapse})]$ and 
$S_2 (\text{collapse})=S_2(0)\exp[-\tilde{D_2}(\text{collapse})]$. 
We quantify the impact of the epidemic 
on populations of each of two species by 
$\Gamma_i$ defined by $\exp(-\Gamma_i)=S_1(\text{collapse})/S_1(0)$. 

Since at the end of the epidemic the number of infected individuals is equal to zero, the overall 
death tolls are given by 
$D_1(\text{collapse})=S_1(0)-S_1 (\text{collapse})$ and 
$D_2 (\text{collapse})=S_2(0)-S_2(\text{collapse})$. 
The fractions of two populations that died during the epidemic
$\rho_1=D_1(\text{collapse})/S_1(0)=1-S_1 (\text{collapse})/S_1(0)=1-\exp(-\Gamma_1)$ 
and $\rho_2=D_2(\text{collapse})/S_2(0)=1-S_2(\text{collapse})/S_2(0)=1-\exp(-\Gamma_2)$ 
are then self-consistently determined by
\begin{eqnarray}
\nonumber
\Gamma_1 & = & (\beta_{11} S_1 (0) /\gamma_1) \cdot  \rho_1 +
(\beta_{12} S_2(0)/\gamma_2) \cdot \rho_2\\ 
\Gamma_2 & = & (\beta_{21} S_1 (0) /\gamma_1) \cdot  \rho_1 +
(\beta_{22} S_2(0)/\gamma_2) \cdot  \rho_2
\label{matrix_collapse}
\end{eqnarray} 
This non-linear system of equations can be numerically (e.g. iteratively) solved for 
$\rho_i=1-\exp(-\Gamma_1)$. 
The solution is fully determined by 
the {\it collapse matrix}:
\begin{equation}
\hat{C}=
\begin{pmatrix}
\frac{\beta_{11}S_{1}(0)}{\gamma_{1}} & \frac{\beta_{12}S_{2}(0)}{\gamma_{2}} \\    
\frac{\beta_{21}S_{1}(0)}{\gamma_{1}} & \frac{\beta_{22}S_{2}(0)}{\gamma_{2}}   
\end{pmatrix}
\end{equation}
Note, that this collapse matrix, describing the cumulative {\it aftermath of an 
epidemic} is subtly yet critically different from the commonly used 
"next generation matrix" \cite{diekmann1990,diekmann2010} describing the dynamics 
at the very {\it start of the epidemic}: 
\begin{equation}
\hat{K}=
\begin{pmatrix}
\frac{\beta_{11}S_{1}(0)}{\gamma_{1}} & \frac{\beta_{12}S_{1}(0)}{\gamma_{2}} \\    
\frac{\beta_{21}S_{2}(0)}{\gamma_{1}} & \frac{\beta_{22}S_{2}(0)}{\gamma_{2}}   
\end{pmatrix}
\nonumber
\end{equation}
One can show that a non-zero collapse with $\Gamma_i>0$ in any of the species is possible 
if and only if the largest eigenvalue of the matrix $\hat{C}$ exceeds 1. 
This does not contradict the classic result \cite{diekmann1990,dobson2004,diekmann2010} 
that the basic reproduction number of the epidemic, $R_{0}>1$, is equal to 
the largest eigenvalue of the next generation matrix $\hat{K}$. The agreement is ensured by the mathematical 
fact that the collapse and the next generation matrices are connected to each other 
by the similarity transformation $\hat{C}=\hat{S}\hat{K}\hat{S}^{-1}$ 
and thus have identical eigenvalues. Here $\hat{S}=S_i(0) \cdot \delta_{ij}$ is 
the diagonal matrix of initial species abundances.
}

{\color{black} In the limit where population collapses 
in both species are large ($\rho_1\sim 1$ and $\rho_2\sim 1$), 
the Eqs. \ref{matrix_collapse} predict
the logarithm of collapse ratios in each of two populations
to be given by a simple sum of matrix elements of the 
collapse matrix: $\Gamma_1=C_{11}+C_{12}$ and $\Gamma_2=C_{21}+C_{22}$. 
In other words, the overall fraction of survivors $\exp(-\Gamma_i)$ is given
by a product of survival probabilities in infections transmitted by the 
members of its own species and those of the opposite species.
This is illustrated by the case of unidirectional transmission shown in Fig. 1, 
where the ``downstream'' species 1 collapses by a factor  
$\exp(-10)=\exp(-5) \cdot exp(-5)=\exp(-C_{11}) \cdot \exp(-C_{12})$, 
while the ``upstream'' species 2 collapses 
only by a factor $\exp(-5)=\exp(-C_{22})$. If the disease was able to spread equally in both directions, both species would suffer equally large collapses $\sim \exp(-10)$.}

{\color{black} Figs. \ref{fig2}-\ref{fig3} show the decimal logarithm }
(as opposed to the natural one) of the species 1 collapse ratio 
$\log_{10}(S_1(0)/S_1(\text{collapse})) =- \Gamma_1/\ln(10)$ for different combinations
of parameters.
\begin{figure}
\includegraphics[angle=0,width=1.0\columnwidth]{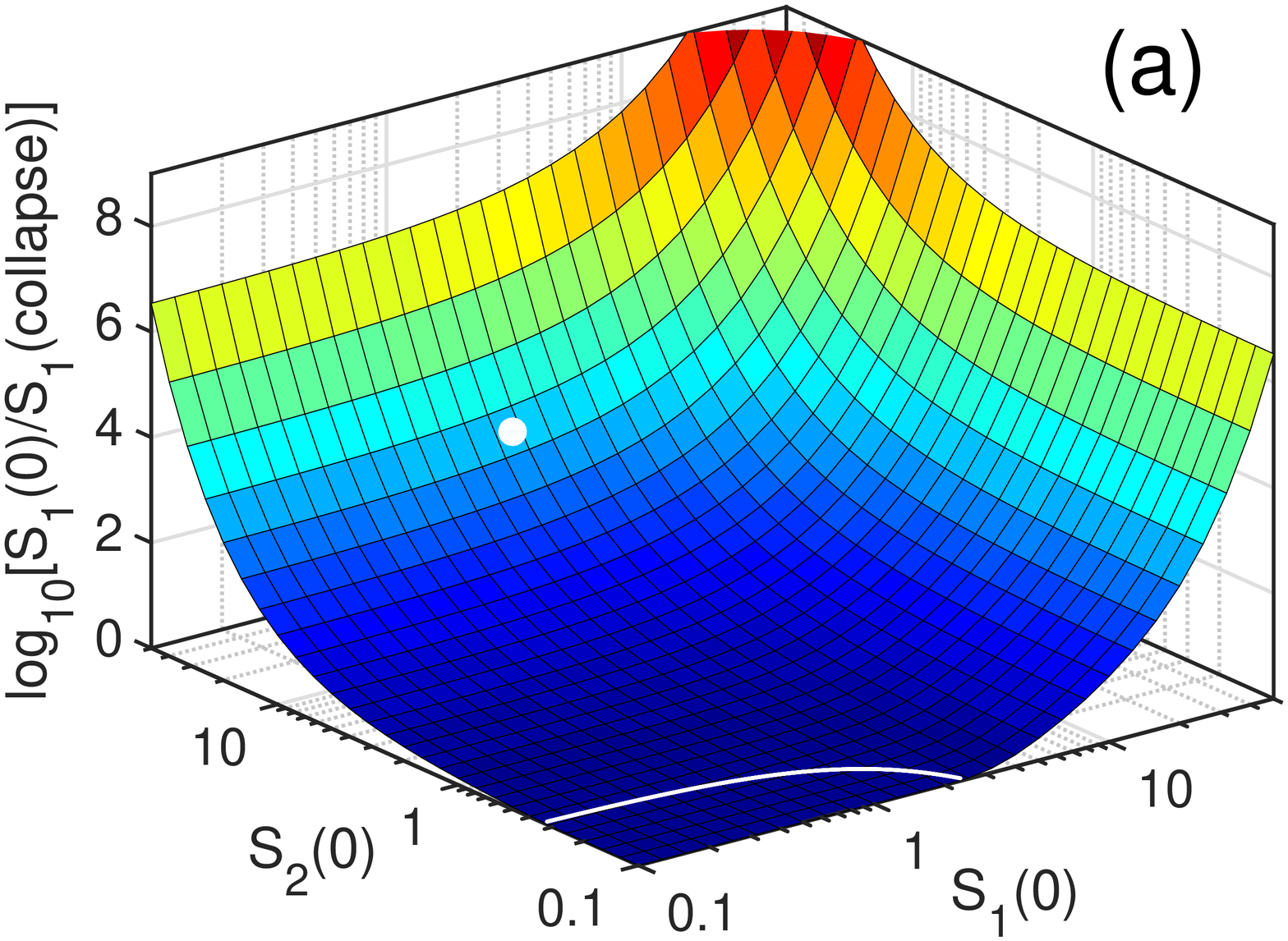}\\
\includegraphics[angle=0,width=1.0\columnwidth]{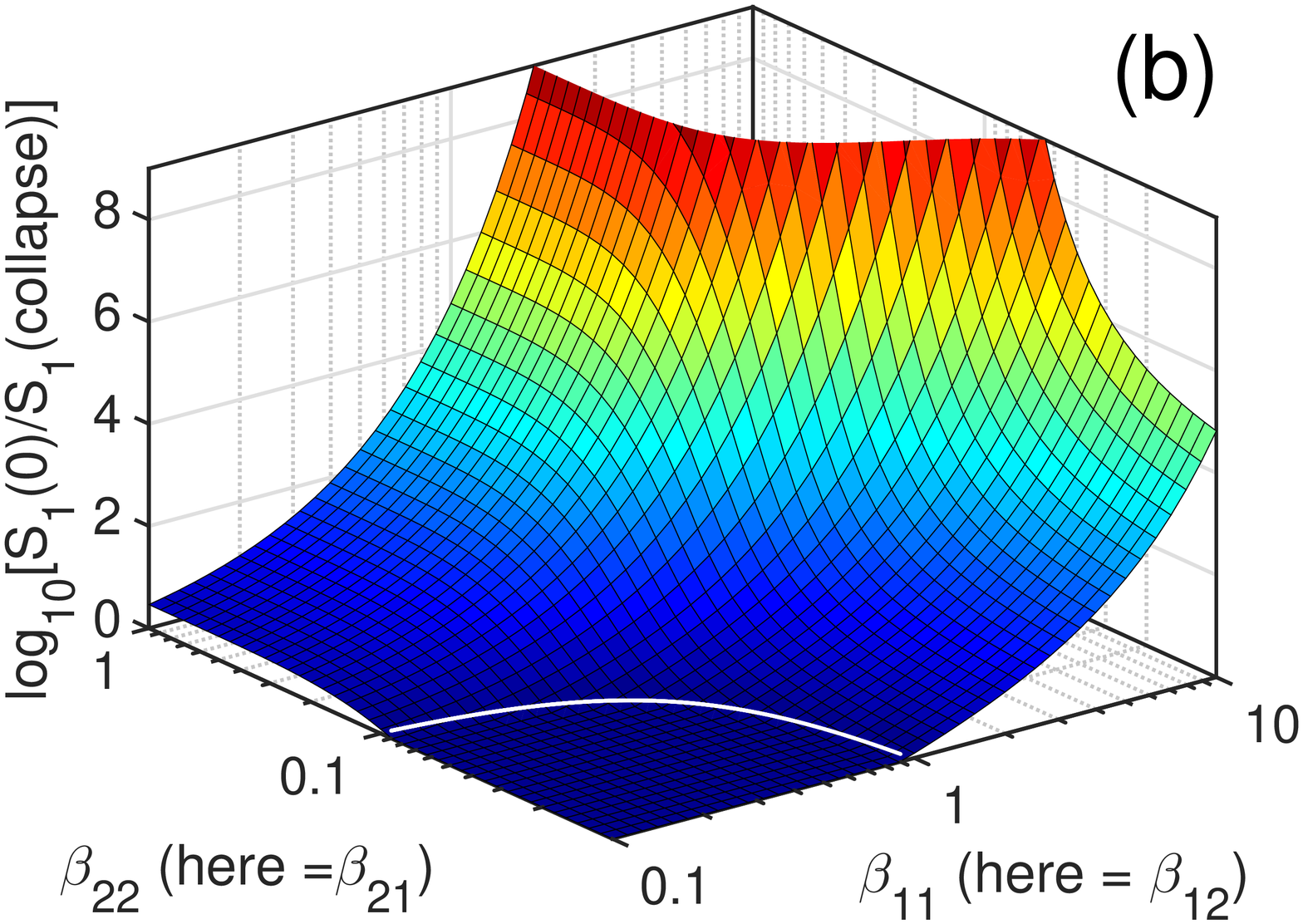}
\caption{Decimal logarithm of the collapse ratio
$\log_{10}(S_1(0)/S_1(\text{collapse}))=\Gamma_1/\ln(10)$
in the population 1 as a function of the two species population sizes (panel (a)) and 
infections rates (panel (b)).
Panel (a) shows the collapse ratio 
as a function of the initial populations $S_1(0)$ and $S_2(0)$ 
in a system where $\gamma_1=\gamma_2=1$ and $\beta_{11}=\beta_{12}=0.3$, 
and $\beta_{22}=\beta_{21}=3$. White line is the predicted epidemic threshold 
at which the largest eigenvalue of the collapse matrix $\hat{C}$ 
is equal to 1. 
Yellow-to-red colors indicate likely extinction of the species 1 with the initial 
population of $10^5$ susceptible individuals.
White dot marks initial population sizes $S_1(0)=1$ and $S_2(0)=10$ used in panel (b), which shows the 
decimal logarithm of the collapse ratio calculated for these initial population sizes and variable infection rates 
$\beta_{11}=\beta_{12}$, and $\beta_{22}=\beta_{21}$. White line marks the predicted epidemic 
threshold.  
}
\label{fig2}
\end{figure}
In Fig. \ref{fig2} we examine the logarithmic magnitude of the species 1 collapse, 
as a function of initial susceptible population sizes of both species (panel (a)) and 
(cross)infections rates (panel (b)). 
Panel (a) plots $\log_{10}(S_1(0)/S_1(\text{collapse}))$ as a 
function of the initial populations $S_1(0)$ and $S_2(0)$ 
in a system where $\gamma_1=\gamma_2=1$ and $\beta_{11}=\beta_{12}=0.3$, 
and $\beta_{22}=\beta_{21}=3$.
White line is the predicted epidemic threshold  below which the largest 
eigenvalue of the collapse matrix $\hat{C}$ 
falls below 1. 
White dot marks the population sizes $S_1(0)=1$ and $S_2(0)=10$ used in the panel (b), which 
shows $\log_{10}(S_1(0)/S_1(\text{collapse}))$ at these population sizes and variable infection rates 
$\beta_{11}=\beta_{12}$, and $\beta_{22}=\beta_{21}$. 
Note that Figs. 2 and 3 shows the decimal logarithm of the collapse ratio. 
{\color{black} Thus 
for a} 
population of, for example, $10^5$ individuals, a collapse 
value greater than 5 (yellow-to-red colors in our Figures 2 and 3)
{\color{black} indicates a likely local extinction threshold for species 1 defined by 
the epidemic reducing the population to (on average) $<1$ surviving individuals.}

In general, {\color{black} two host species in our model are } characterized by different infection parameters
and potentially highly asymmetric transmission rates. For example, for Ebola virus in bats and gorillas mentioned 
above \cite{bermejo2006,leroy2005} cross infections are believed to be mediated primarily by bats' droppings landing 
on fruits that gorillas eat. Thus the spread of the virus is generally uni-directional from 
species 2 (bats) to species 1 (gorillas). In Fig. \ref{fig0} we simulated our model with $\beta_{21}=0$.  
In Fig. \ref{fig2} we examine how the magnitude of the post-epidemic drop in population sizes depends 
on parameters.  Fig. \ref{fig2}a shows the dependence of 
the logarithmic collapse ratio in the species 1 (gorillas) on the size of
the magnitude of cross-species collapse matrix element 
$C_{12}=\beta_{12} S_2(0)/\gamma_2$ and the size of 
intra-species collapse number $C_{22}$ within the 
species 2 (bat) population.
To further illustrate our point we selected the basic collapse number 
in the population of  gorillas to be well below the species 1  
epidemic threshold if it was isolated from species 2 ($C_{11}=0.1<<1$). 
Yet, we were anyway able to get an extinction-level collapse in the population of ``gorillas'' 
as long as the majority of bats were infected. It is important to note that 
our model equally well applies to the case where species 2 (bats) do not die in 
the course of the epidemic but are instead removed from the ranks of susceptible 
population by becoming immune to the disease 
{\color{black}
(see the discussion for generalization of our mathematical formalism to 
incorporate recovery with immunity).} The species 2 death 
rate $\gamma_2$ in this case 
is simply the rate at which they acquire immunity and thus stop being infectious.
\begin{figure}
\includegraphics[angle=0,width=1.0\columnwidth]{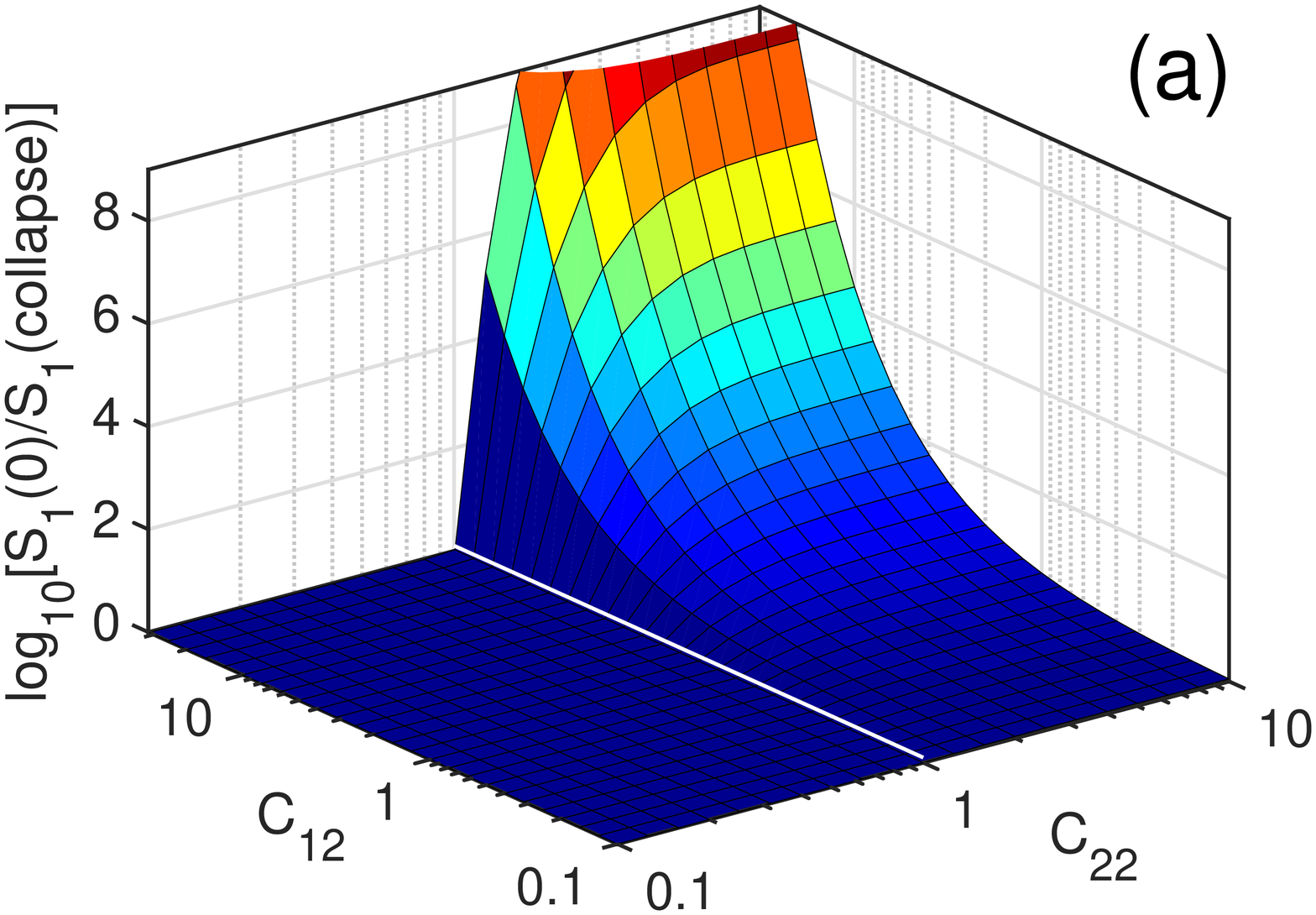}\\
\includegraphics[angle=0,width=1.0\columnwidth]{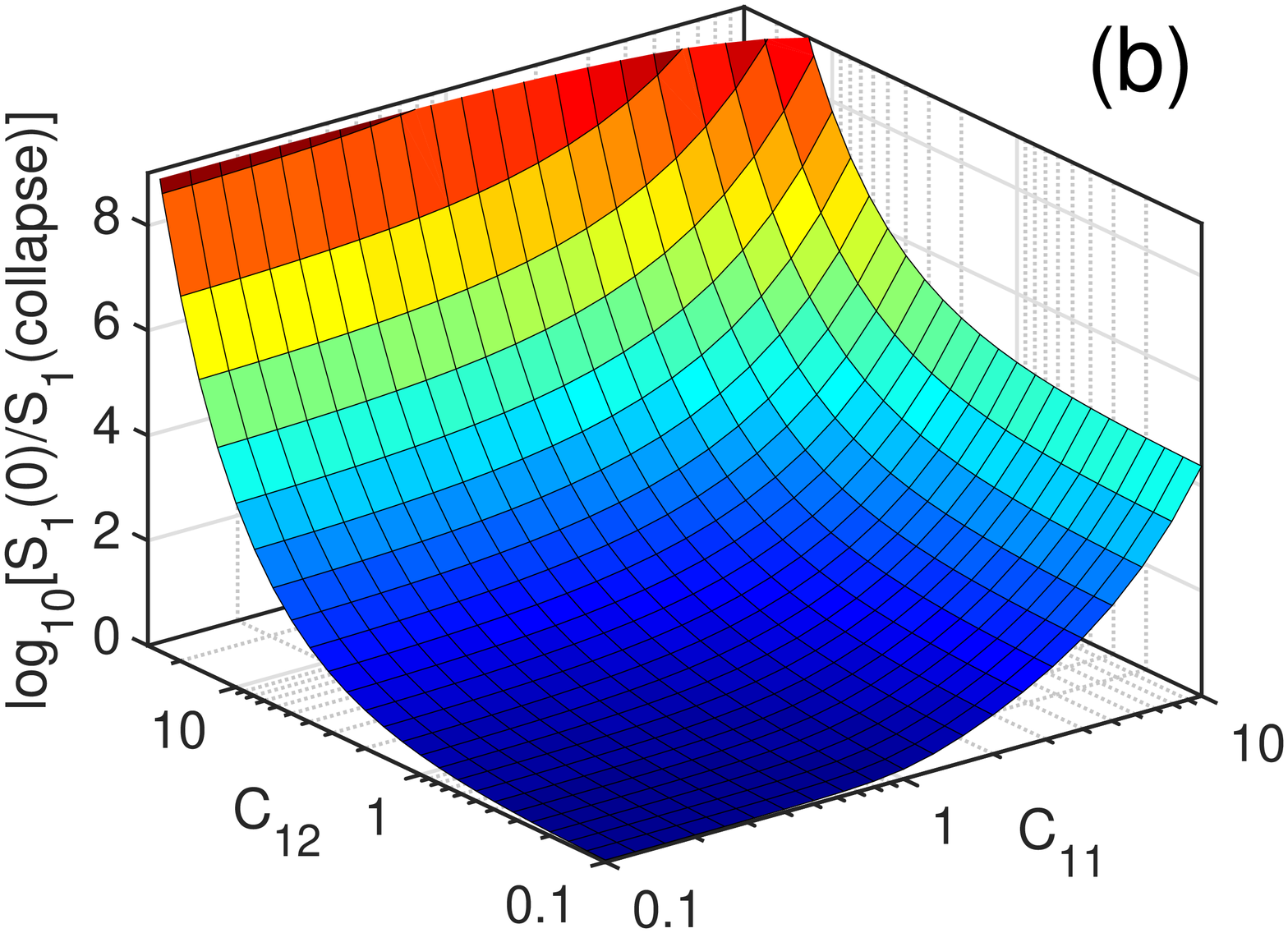}\\
\caption{Collapse ratio $S_1(0)/S_1(\text{collapse}$ of the 
population of the species 1 in the case of uni-directional 
transmission: $C_{21}=0$, following an epidemic  
started with a very small number of infected
individuals ($I_1(0)=I_2(0)=10^{-6}$).
(a) Collapse ratio in the population of the  species 1 
with fixed intra-species {\color{black} collapse factor} 
$C_{11}=0.1$ as a function 
of the species 2 collapse number  $C_{22}$ and cross-species 
collapse number $C_{12}$ quantifying disease transmission from species 2 to 1.
White line is the predicted epidemic threshold below which the overall 
reproduction number falls below 1. 
(b) Collapse ratio 
the population of species 1, with a fixed value of $C_{22}=2$ and variable 
$C_{11}$ and $C_{12}$. There is no epidemic threshold in this case 
as the basic reproduction number in the species 2 is selected to be 
larger than 1 so that the epidemic would always be able to start.
}
\label{fig3}
\end{figure}

The properties of the system can be further analyzed in terms
of a simple analytic expression obtained in the limit where 
$\Gamma_1 \gg 1$ and  $\Gamma_2 \gg 1$ so that $\rho_1 \simeq 1$ and 
$\rho_2 \simeq 1$ 
{\color{black} (strictly speaking this is the limit of the model where  
$\Gamma_i , \; \Gamma_2 \rightarrow \infty$)}. 
In this case the Eq. \ref{succ} becomes simply 
\begin{eqnarray}
\log(\frac{S_1(0)}{S_1(\text{collapse})}) &=& \Gamma_1
= C_{11}+C_{12}
\label{collapse1}\\
\nonumber
& = &
\frac{\beta_{11} S_1(0)}{\gamma_1} +
\frac{\beta_{12} S_2(0)}{\gamma_2} 
\end{eqnarray}
This limit approximately describes the simulations shown in  
Fig. \ref{fig0} where the collapse of the first species is very close to  
$S_1(\text{collapse}) \simeq \exp(-C_{11}-C_{22})=\exp(-10)$ (see blue arrow).
Note that the population collapse in the species 1 
described by the Eq. \ref{collapse1} does not depend on the impact of the epidemic on 
the species 2 population,
corresponding to a near complete elimination of the susceptible population 2 ($\Gamma_2 \gg 1$).
This limit can be seen in Fig. \ref{fig2}a as leveling off of the surviving fraction of the species 1 
for large values $C_{22} \gg 1$, corresponding to saturation of 
the reservoir of the species 2. Fig. \ref{fig2}b further 
explores this limit by plotting $\Gamma_1$ as a function of $C_{11}$ and $C_{12}$ for 
a fixed ``bat-to-bat'' (within-species 2) collapse {\color{black} factor} 
$C_{22}=2$.
In this case a large fraction of the population 2 ($1-\exp(-2)$ or 86\%) becomes infected thus 
opening up plentiful opportunities (broad range of two other parameters of the model) for an 
extinction-level collapse of the population 1.

\section*{Discussion}
Diseases are a real and constant danger for nearly any of the 
species on our planet, and are occasionally 
assumed to drive or facilitate 
extinction-scale events \cite{lyons2004,wake2008,fisher2012}.
This paper demonstrated that such events would be more likely 
when a lethal pathogen infects more than one host  species .  
Above we explored a simple two species model 
subject to epidemic-driven
population collapses and extinctions. 
The epidemic could be triggered by 
either the appearance of a new pathogen or a 
sudden increase in intra- or cross-species infection rates 
in a new ecological layout. 
{\color{black} As can be inferred from the Eq. \ref{collapse1} 
a severe population collapse of the species 1 is favored by an initially 
large population of the co-infecting species 2 
(large $S_2(0)$) that can stay infectious for a 
long time ($\gamma_2$ small) resulting in a large 
cross-species collapse number $C_{12}$.}
{\color{black}. Cross-species transmission could dramatically 
amplify the collapse due to within-species transmission which 
could even be characterized by a sub-critical value of $R_{0}(1 \to 1)<1$
($C_{11}$ in our notation). 

If a population would survive the first epidemic,
one may speculate whether it would be sustainable in 
the long term endemic steady state. This was previously
considered by \cite{dobson2004}, with the overall
result was that coexistence of two or more species in the 
endemic steady state depends on multiple species-specific
parameters. 
According to Ref. \cite{dobson2004}, the extinction of species 
in the endemic state is possible when intra-species transmission
is high and it targets host species in the inverse order of their 
growth rates.  
That is to say, slowly growing species will go extinct first 
when they share pathogens with faster growing ones. 
Thus species survival in the endemic state of the disease depends on different parameters 
(growth rates) than in the initial epidemics (relative population sizes). 
}

Our study suggests that 
transient  epidemics of diseases provide 
species with powerful ``weapons" against each other.
Such weapons have been well documented in 
the microbial world where bacterial species co-infected by 
the same phage 
\cite{lederberg1957} 
fight ongoing battles with each other and their 
phage pathogen. 
Long history of such "red-queen" 
evolutionary dynamics can be inferred from 
many-layered defense and counter-defense mechanisms 
encoded within their genomes \cite{labrie2010}.
The use of shared diseases as a weapon
have similarity to the apparent competition
between multiple prey species sharing a common predator
\cite{holt1977}. 
Our analysis extends these earlier results by including 
the impact of transient epidemics, and adding the possibility that
 permanently remove an otherwise fit predator.

An important example of cross-species interactions occurs
when a single pathogen co-infects a highly abundant 
prey and its typically much its less abundant 
predator. 
Mapping the prey to 
species 2 in our model 
this situation would give rise to particularly large values
of $C_{22}$ and $C_{12}$ which are both proportional 
to prey's high population density $S_2(0)$. 
Our results suggest that 
such a 
disease may only need
to be present during a relatively short period,
in order to locally eliminate the less abundant predator species.
Such disease would also result in a short-term population 
loss of the prey, but give it a long term gain 
in terms of eliminating the predator entirely.
Alternatively, for the pathogen 
it would be
evolutionary beneficial to be either completely 
benign or at least less deadly 
to its prey host, but much more lethal to its host's predators.
Indeed, by reducing predator population it increases prey (and hence its own) 
population.
Thus in contrast to the classical single host 
results of \cite{dubos1965,may1990},
our analysis suggests that it is not always beneficial 
for a disease to become more benign to {\it all of its hosts.} 

{\color{black} Diseases often leave a substantial fraction of survivors, and their 
epidemics only cause a finite-size collapse in populations of their hosts.
Somewhat counterintuitively this may increase the diversity of the host ecosystem 
by allowing hosts to bypass the competitive exclusion principle, according to which 
only the single fastest growing species survives in the long run. 
One example we investigated before \cite{maslov2016} is 
the negative density-dependent selection in which 
phage epidemics preferentially spread in bacterial species or strains
with large populations (so called "Kill-the-Winner" principle 
\cite{thingstad2000}) thereby leading to their abrupt and 
severe collapse.

Since the focus of this study is on extinction level population collapses, 
above we considered an extreme case of a disease with 100\% mortality.
Yet our results can be
readily extended to a more general case in which 
a fixed fraction $x_i$ of infected individuals of species $i$ die, while 
$1-x_i$ - recover with full immunity. As was discussed above,  
for the purposes of the SIR mathematical model without birth 
these two outcomes are identical. 
Let $\gamma_i$ denote the overall rate of death and recovery with immunity. 
Out of a fraction $1-\exp(-\Gamma_i)$ removed from the corresponding 
susceptible population, $x_i(1-\exp(-\Gamma_i))$ actually died, 
while $(1-x_i)(1-\exp(-\Gamma_i))$ survived. 
Thus by the end of the first epidemic the overall (both susceptible and immune) 
surviving population fraction 
is given by $1-x_i+x_i\exp(-\Gamma_i)$, where as before 
$\Gamma_i$ is determined by the Eq. \ref{matrix_collapse}. 
Coming back to the bats and gorillas example considered above 
one can have a situation in which the Ebola 
virus is rather deadly ($x_1 \simeq 1$) for one of the species (gorillas), 
while being mild in another ($x_1 \simeq 0$) (bats \cite{leroy2005}). 
In this case the severe collapse of the gorilla population 
continues to be described by the Eq. \ref{collapse1}. 

In spite of its simplified well-mixed mass-action kinetics, our results suggest a way on how to minimize the probability of a 
disastrous collapse in human populations. {\color{black} Humans 
routinely share pathogens with animals. Indeed, }
more than half of nearly 1500 known human pathogens are shared with 
at least one animal host \cite{woolhouse2005}.
Wolfe et al. \cite{wolfe2007} classified such zoonotic diseases into 
5 categories (called evolutionary stages) out of which 
stages 2-4 differ from each other exclusively by their 
basic reproduction number in human-to-human transmission  
($C_{11}$ in our notation). Stage 2 is characterized by 
a complete lack of human-to-human transmission ($C_{11}=0$), 
Stage 3 - by sub-critical human-to-human transmission $0<C_{11}<1$), 
and Stage 4 - by super-critical human-to-human transmission ($C_{11}>1$).
Human-to-human basic reproduction number, $C_{11}$, is clearly important both 
for endemic state stability considered 
in Ref. \cite{dobson2004} as well as for the epidemic-driven population 
collapse considered here, especially in the 
case where the disease does not spread on its own in their animal host ($C_{22}<1$).
However, as demonstrated in Fig. \ref{fig2}b, for a pathogen capable to spread in
{\color{black} the co-infected animal ($C_{22}>1$ such as example used in Fig. \ref{fig2}b)} 
the human-to-human basic reproduction number $C_{11}$ has only mild and 
qualitative impact on $\Gamma_1$ quantifying the logarithm of the population collapse 
in humans. Much more important factor is the magnitude of the animal-to-human {\color{black}
collapse factor $C_{12}=\beta_{12} S_2(0)/\gamma_2$ }. It is proportional to 
the population of the animal host, which could potentially be very large. 
We are outnumbered by populations of, for example, small birds, rats and mice.
Our paper emphasizes the advantage of limiting our
exposure to such species with large populations and high 
growth rates.
Perhaps much of the recent trend showing 
the overall decrease in occurrence of serious epidemics in the 
industrial world could be attributed to progressively less frequent 
contacts between humans living in major population centers and these 
animals. To prevent serious disease outbreaks in the future 
it may be particularly useful to closely monitor
abundant disease carriers in regions with high potential 
for inter-species contacts.

\section*{Acknowledgments}

We thank Lone Simonsen for 
inspiring discussions and for bringing to our attention 
2002-2003 Ebola epidemics among gorillas.

\end{document}